\documentclass{article}
\usepackage{spconf,amsmath,graphicx,multirow,array}
\usepackage{epstopdf}
\usepackage{amssymb}
\usepackage{caption}
\usepackage{algorithm} 
\usepackage{algorithmic} 
\usepackage{color}
\usepackage{hyperref}
\usepackage{mathrsfs}
\usepackage{booktabs}
\usepackage{cite}
\definecolor{note}{RGB}{105, 105, 105} 
\usepackage{setspace}

\usepackage{subfigure}

\title{Speech Enhancement with Intelligent Neural Homomorphic Synthesis}
\name{
\begin{tabular}{@{}c@{}}
Shulin He\textsuperscript{1,2,$\dagger$}\thanks{$^{\dagger}$ Work done during internship at Tencent Ethereal Audio Lab.}, Wei Rao\textsuperscript{2}, Jinjiang Liu\textsuperscript{1}, Jun Chen\textsuperscript{2}, Yukai Ju\textsuperscript{2}, Xueliang Zhang\textsuperscript{1}, \\ Yannan Wang\textsuperscript{2}, Shidong Shang\textsuperscript{2}
\end{tabular}
}

\address{
\textsuperscript{1}College of Computer Science, Inner Mongolia University, China\\
\textsuperscript{2}Tencent Ethereal Audio Lab, Tencent Corporation, Shenzhen, China\\
\small \texttt{heshulin@mail.imu.edu.cn,cszxl@imu.edu.cn}
}

\begin{document}
\ninept
\setstretch{0.875} 
\maketitle
\begin{abstract}
Most neural network speech enhancement models ignore speech production mathematical models by directly mapping Fourier transform spectrums or waveforms.
In this work, we propose a neural source filter network for speech enhancement.
Specifically, we use homomorphic signal processing and cepstral analysis to obtain noisy speech's excitation and vocal tract.
Unlike traditional signal processing, we use an attentive recurrent network (ARN) model predicted ratio mask to replace the liftering separation function.
Then two convolutional attentive recurrent network (CARN) networks are used to predict the excitation and vocal tract of clean speech, respectively.
The system's output is synthesized from the estimated excitation and vocal.
Experiments prove that our proposed method performs better, with SI-SNR improving by 1.363dB compared to FullSubNet.

\end{abstract}
\begin{keywords}
Speech enhancement, neural network, ARN
\end{keywords}
\section{Introduction}
\label{sec:intro}

Noise is ubiquitous in life and inevitably affects the use of remote calls, human-machine interaction, and hearing aids.
Speech enhancement aims to improve a speech signal's intelligibility and quality by removing or attenuating background noise.
Despite the long history of speech enhancement research, it is still a challenge in real-world scenarios.

Traditional speech enhancement methods include spectral subtraction \cite{boll1979suppression}, Wiener filtering \cite{lim1979enhancement}, and minimum mean-squared-error estimation \cite{ephraim1984speech}.
These algorithms are capable of handling well only stationary noise. Thus, their applications are minimal.
Deep neural networks (DNNs) have made significant progress in speech enhancement tasks since speech enhancement was first considered as a supervised learning problem in \cite{wang2013towards}.

Typically, DNN-based speech enhancement converts the voice signal to a time-frequency representation, which is then used to enhance noisy speech \cite{wang2022attention,liu21f_interspeech,tan2018convolutional,li2022filtering}.
There are two training targets: those based on masking and those based on mapping \cite{wang2018supervised}.
Most T-F representation-based approaches try to improve only magnitudes, whereas time-domain signal reconstruction uses the noisy phase unchanged.
A recent study determined that phase has a crucial role in improving speech quality, particularly in low SNR environments \cite{mowlaee12_interspeech}.
This has prompted academics to investigate ways for improving complex spectrums.
Utilizing complex spectrums increases the effectiveness of speech augmentation.

Recent research \cite{pandey2022self,pandey2019tcnn,pandey2022tparn, li2022use} employs a time-domain strategy to circumvent the phase estimation difficulty by directly recreating waveform samples. In order to have sufficient spectrum resolution, short-time processing based on a T-F representation requires frame size to be greater than a certain threshold, whereas, in time-domain processing, frame size can be set arbitrarily. In \cite{luo2020dual,luo2019conv,ge2021multi}, the performance of a time-domain speaker separation network is significantly enhanced by minimizing the frame size. Using a smaller frame size, however, necessitates additional computations due to the increased number of frames.

Compared with the magnitude spectrums, the complex spectrums and waveforms lack distinct structures, making the task more difficult \cite{du2020self}.
For example, the formant structure of speech is apparent in the magnitude spectrums, but it is not clear in the complex spectrums and waveforms.
In \cite{du2020joint}, the researchers enhanced the Mel spectrums of noisy speech and then used a vocoder to restore the enhanced Mel spectrums to the original waveform.
The experimental results show the effectiveness of these methods, but the computational complexity makes vocoders unsuitable for practical use.

Relying on the recent research on neural homomorphic vocoder with low computational complexity \cite{liu2020neural}, Jiang et al. have studied a speech enhancement algorithm using neural homomorphic synthesis (NHS-SE) \cite{jiang2022speech}. NHS-SE combines the advantages of a DSP-based vocoder and complex-valued neural network based spectrum denoiser, yielding state-of-the-art performance in terms of PESQ and eSTOI. Specifically, a complex-valued homomorphic filtering system is firstly used to decompose the speech signal into the excitation and vocal tract. 
Later, two complex convolutional recurrent networks are adopted for estimating the clean spectrums of the excitation and vocal tract, respectively.
Lastly, the enhanced speech is synthesized with denoised components.
The Excitation and vocal tract have a more explicit structure than complex spectrum and waveform signals.
However, this method has two drawbacks: 1) since the traditional DSP method of homomorphic signal processing uses a simple segmentation function (liftering) to obtain the excitation and sound channel, it does not match the actual situation. Different speakers have different ideal liftering coefficients. And 2) the traditional homomorphic signal processing process does not apply to small discrete Fourier transform (DFT) sizes, which brings greater computational complexity. 

Inspired by the above work, we propose a novel method for speech enhancement with neural homeostatic synthesis.
We use a neural homomorphic signal processing module to process the noisy speech after framing to obtain its excitation and vocal tract.
Unlike traditional homomorphic signal processing methods, we use an attentive recurrent network (ARN) block to predict a mask as the segmentation filter for the excitation and the vocal tract.
Then two convolutional attentive recurrent network (CARN) networks predict the target complex spectrum of excitation and vocal tract, respectively.
The estimated clean speech is synthesized from the estimated excitation and vocal tract.
Experiments show that the proposed speech enhancement method outperforms baselines.
Compared to the neural homomorphic synthesis method proposed in \cite{jiang2022speech}, SI-SNR is improved by 0.657 dB in the reverberant environment.
Furthermore, for the small DFT size, the method in \cite{jiang2022speech} cannot separate the excitation and vocal tract due to the use of traditional homomorphic signal processing leading to severe degradation of the metrics.
In contrast, our proposed method has no significant degradation.

\section{Intelligent neural homomorphic synthesis}
\label{sec:Intelligent neural homomorphic synthesis}
\begin{figure}[ht]
	\centering
	\includegraphics[height=2.4cm]{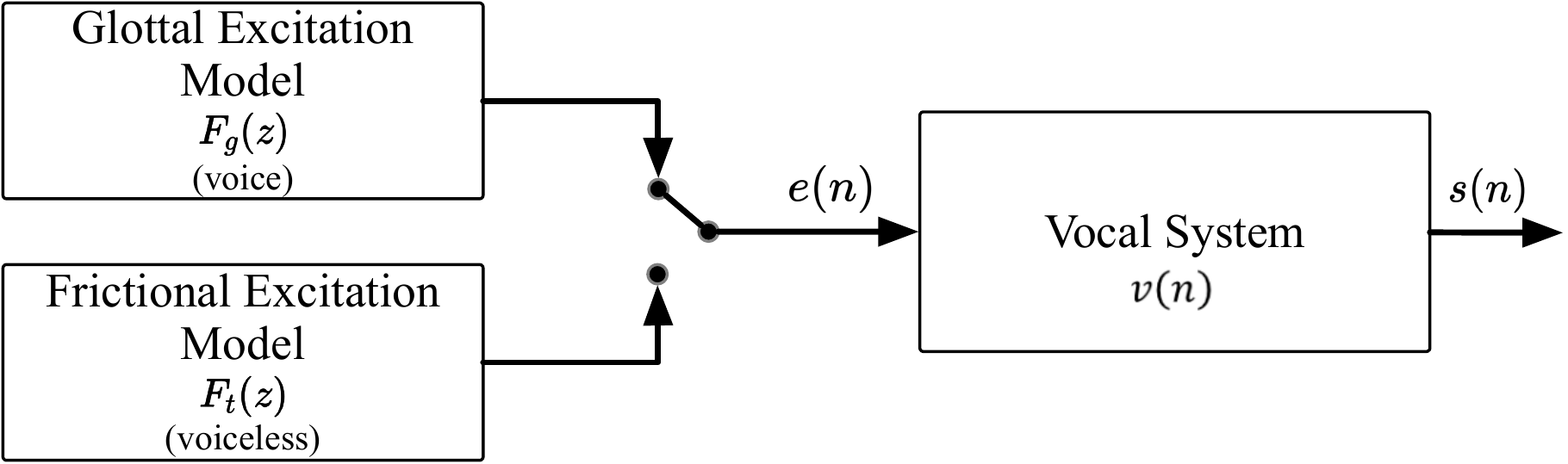}
	\caption{Block diagram of a simplified source-filter model.}
	\label{fig:sourcefilter}
\end{figure}
\vspace{-0.4cm}
\subsection{Source‐Filter Model}
\label{ssec:Source‐Filter Model}
In traditional signal processing, digital models are often used to quantitatively describe the process of speech signal generation, with the source-filter model being the most widely used \cite{rabiner2010theory}. The source-filter model includes the excitation model, the vocal tract model, and the radiation model, which correspond to the excitation formed by the joint action of the airflow and the vocal folds, the tuning movement of the vocal tract, and the radiation effect of the lips and nostrils, respectively. The vocal tract and radiation model are usually referred to as the vocal system for concise analysis. The relationship between them can be represented in Figure \ref{fig:sourcefilter}.

Two types of these excitations $e[n]$ are voice and voiceless.
When a voice is produced, the airflow strikes the taut vocal folds to produce vibrations, causing a quasi-periodic pulse train to form at the voice gate and use it to excite the vocal tract.
When a voiceless is produced, the vocal folds are relaxed without vibrations, and the airflow enters the vocal tract directly through the vocal folds.
The vocal tract system $v(n)$ represents the effect of vocal tract resonance and lip radiation.
The speech signal $s(n)$ is the output of the vocal system $v(n)$ excited by excitation signal $e(n)$, i.e., $s(n) = e(n) \ast v(n)$.

\subsection{Traditional homomorphic signal processing}
\label{ssec:Traditional homomorphic signal processing}

According to the source-filter model theory, the generation process of the speech signal is as follows:
\begin{equation}
	s(n)=e(n) * v(n).
\end{equation}
where $s(t)$, $e(t)$, and $v(n)$ denote the speech, excitation, and vocal tract signals with the time index of $n$, respectively. Homomorphic signal processing can decompose two convolutional signals, consisting of three steps. First, DFT is applied to the framed speech, which transforms the convolutional signal into a multiplicative signal, at which point the spectrum of the input signal is obtained as follows:
\begin{equation}
 \begin{split}
	&\operatorname{DFT}[s(n)]=S(z)\\
	&S(z) =E(z) \times V(z)
	\end{split}
\end{equation}
where $z$ is the frequency index. The frequency domain speech is obtained by multiplying the excitation and the vocal tract in the frequency domain.

Second, perform logarithmic operations on frequency-domain speech $S(z)$ to turn the product signal into an addition signal:
\begin{equation}
	\log S(z)=\log E(z)+\log V(z).
\end{equation}
The current speech, excitation, and vocal tract signals are defined as $\hat{S}(z)$, $\hat{E}(z)$, and $\hat{V}(z)$, respectively.

Since this signal is an additive logarithmic spectrum, which is not convenient to use, a common approach is to change it back to the time domain for processing. So the third part performs the inverse DFT (IDFT) operation, and what we get is the cepstrum of the speech signal: 
\begin{equation}
    \operatorname{DFT}^{-1}(\hat{S}(z))=\operatorname{DFT}^{-1}(\hat{E}(z)+\hat{V}(z)).
\end{equation}
In the cepstrum domain, the relationship between speech $\hat{s}(n)$, excitation $\hat{e}(n)$, and vocal tract $\hat{v}(n)$ is expressed as follows:
\begin{equation}
    	\hat{s}(n)=\hat{e}(n) * \hat{v}(n).
\end{equation}

Then the time-domain cepstrum signal is decomposed into excitation and vocal tract by cepstrum liftering. Vocal tract liftering is defined as follows:
\begin{equation}
l_v(n)=\left\{\begin{array}{ll}
1, & \mathrm{n}<\mathrm{N} \\
0, & \mathrm{n} \geq \mathrm{N}
\end{array}\right.
\end{equation}
To obtain the excitation information, its corresponding representation of liftering is as follows:
\begin{equation}
l_e(n)=\left\{\begin{array}{ll}
1, & \mathrm{n} \geq \mathrm{N} \\
0, & \mathrm{n} <  \mathrm{N}
\end{array}\right.
\end{equation}
where $N$ is the quefrency for sperating the cepstrum, set to 29.
\vspace{-0.4cm}
\subsection{Intelligent liftering}
\label{ssec:Intelligent liftering}
The difficulty in traditional signal processing is how to determine the parameter $N$ for liftering. Usually, liftering uses statistics to determine the value of $N$. However, it is known that statistics are not accurate. Moreover, since the excitation and vocal tract are not only one of them on each time-frequency (T-F) bin, the traditional segmentation function liftering does not perfectly split the excitation and vocal tract. To solve this problem, we propose a novel intelligent liftering. The intelligent liftering estimate a ratio mask split excitation and vocal tract in the range of 0 to 1 through an ARN module $\mathscr{A}_l$, as follows:
\begin{equation}
    \begin{split}
    &M = \operatorname{SIGMOID}(\mathscr{A}_l(\hat{s}(n)))\\
    &\hat{e}(n) = M \times \hat{s}(n)\\
    &\hat{v}(n) = (1-M) \times \hat{s}(n)
    \end{split}
\end{equation}
In the cepstrum domain, the dynamic range of the excitation and the vocal tract differ significantly, so intelligent liftering can then distinguish between the two in the unsupervised case. Unsupervised intelligent liftering cannot determine the order of excitation and vocal tract. However, the order of the excitation and vocal tract does not affect the performance of speech enhancement with neural homomorphic synthesis.
\vspace{-0.4cm}

\section{Intelligent Neural Homomorphic Synthesis}
\label{sec:Intelligent Neural Homomorphic Synthesis}
\subsection{Architecture}
\label{ssec:Architecture}
\begin{figure*}[ht]
	\centering
	\includegraphics[height=3.5cm]{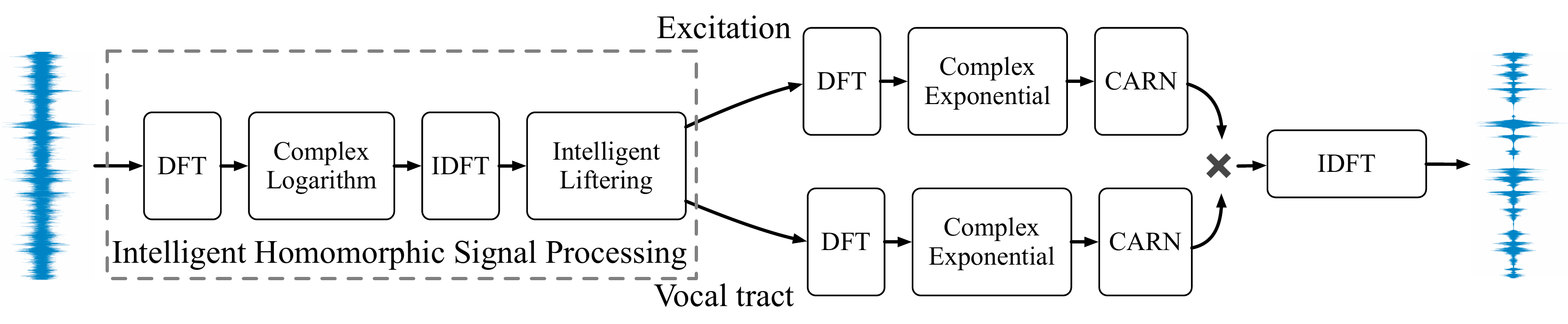}
	\caption{Overview of the proposed system.}
	\label{fig:overview}
\end{figure*}
Figure \ref{fig:overview} shows the block diagram of the proposed method. First, we obtain the excitation and vocal tract of noisy speech by the proposed intelligent homomorphic signal processing. Intelligent homomorphic signal processing includes DFT, logarithm, IDFT, and intelligent liftering. Secondly, the exponents are calculated after doing DFT on excitation and vocal tract to get the representation of excitation and vocal tract in the Fourier domain. The third step models the two spectrums using two CARNs with the same structure. Finally, the estimated clean speech spectrum is obtained by multiplying the two spectrums.
The whole process can be expressed as follows:
\begin{equation}
 \begin{split}
	&\hat{e}_l(n), \hat{v}_l(n)= M \times \hat{s}(n),(1-M) \times \hat{s}(n)\\
&\tilde{V}_l(z) = \mathscr{W}_e\left(\exp \left(\operatorname{DFT}\left(\hat{v}_l(n)\right)\right)\right) \\
&\tilde{E}_l(z) = \mathscr{W}_v\left(\exp \left(\operatorname{DFT}\left(\hat{e}_l(n)\right)\right)\right)\\
&\tilde{s}(n)=\operatorname{DFT}^{-1}\left(\tilde{E}_l(z) \times \tilde{V}_l(z)\right)
	\end{split}
\end{equation}
where $\mathscr{W}_e$ and $\mathscr{W}_v$ are CARNs for estimating the target excitation and vocal tract, respectively. 

The proposed system has intelligent liftering and two CARNs with trainable parameters. All operations are derivable

\label{ssec:CARN}
\begin{figure}[ht]
	\centering
	\includegraphics[height=4.7cm]{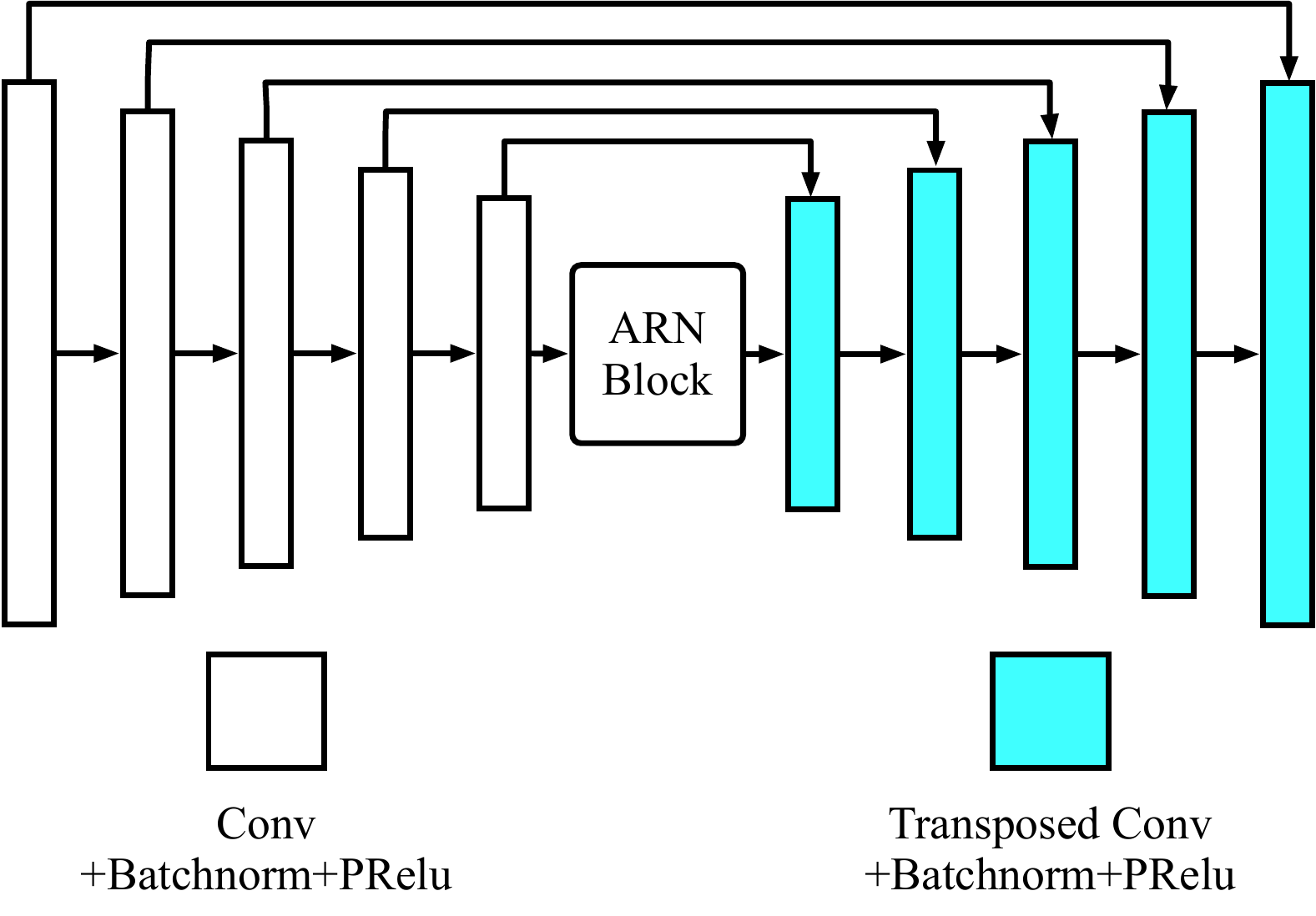}
	\caption{Structure of CARN.}
	\label{fig:carn}
\end{figure}
\subsection{CARN}
\begin{figure}[ht]
	\centering
	\includegraphics[height=2cm]{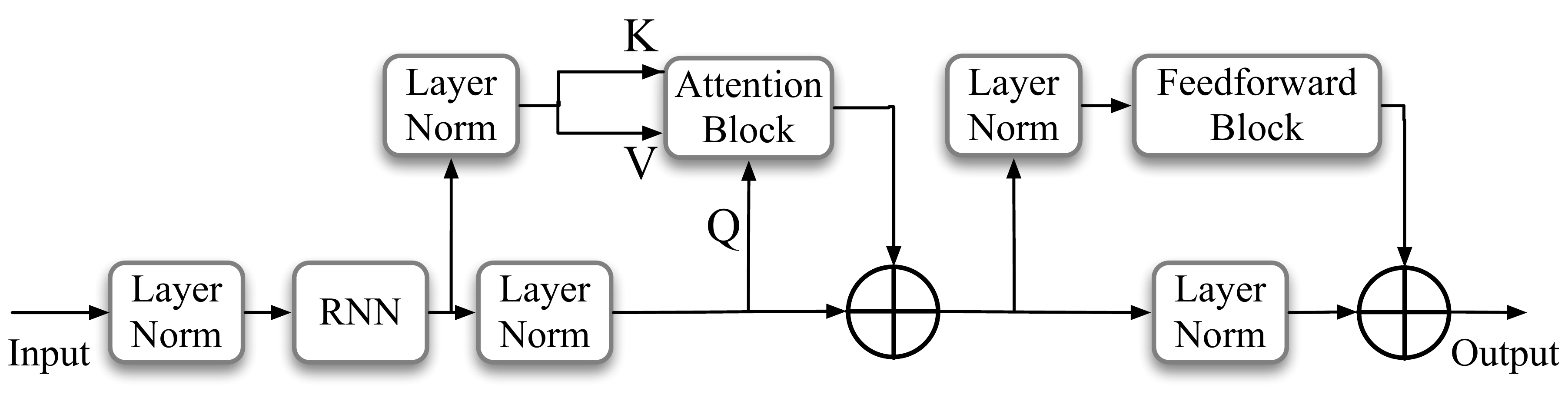}
	\caption{Structure of ARN.}
	\label{fig:arn}
\end{figure}

We use the CARN as the key component for complex excitation and vocal tract spectrum mapping-based speech enhancement, as shown in Figure \ref{fig:carn}. The network architecture is based on CRN, which models temporal dependencies using a convolutional encoder-decoder structure and a recurrent neural network (RNN) bottleneck. Such an architecture successfully captures a given input's local and global contexts. We combine the real and imaginary components of the complex excitation or vocal tract and send 3-D feature maps to the CARN. The CARN encoder is a convolutional neural network (CNN) downsampler that reduces the feature dimension along the frequency axis using convolutions. The decoder mirrors the encoder architecture to restore the feature dimension with transposed convolutions. As shown in Figure \ref{fig:arn}, we replace the RNN of the CRN with the ARN module \cite{pandey2022self}.

\subsection{LOSS FUNCTION}
\label{ssec:LOSS FUNCTION}
The training objective of the proposed system consists of two parts.
First, we apply a scale-invariant signal-to-noise ratio (SI-SNR) \cite{luo2019conv} loss, which is a time domain loss function as follows:
\begin{equation}
\begin{array}{l}
\mathbf{s}_{\text {target }}=\frac{\langle\hat{\mathbf{s}}, \mathbf{s}\rangle \mathbf{s}}{\|\mathbf{s}\|^2} \\
\mathbf{e}_{\text {noise }}=\hat{\mathbf{s}}-\mathbf{s}_{\text {target }} \\
\mathcal{L}_{\text {si-snr }}=10 \log _{10} \frac{\left\|\mathbf{s}_{\text {target }}\right\|^2}{\left\|\mathbf{e}_{\text {noise }}\right\|^2},
\end{array}
\end{equation}
where $\hat{\mathbf{s}} \in \mathbb{R}^{1 \times T}$ and $\mathbf{s} \in \mathbb{R}^{1 \times T}$ refer to the estimated and original clean sources, respectively, and $\|\mathbf{s}\|^2=\langle\mathbf{s}, \mathbf{s}\rangle$ denotes the signal power.

\begin{table*}[!htb]
\small
  \centering
  \caption{The performance of SI-SNR, STOI, WB-PESQ, and NB-PESQ on the DNS Challenge dataset. (320, 160) (512, 256) means that the window length and frame shift of the Fourier transform in this method are 320, 160 or 512, 256 points, respectively.}
    \begin{tabular}{ccccccccc}
    \toprule
    \multirow{2}[4]{*}{Model} & \multicolumn{4}{c}{With Reverb} & \multicolumn{4}{c}{Without Reverb} \\
    \cmidrule(lr){2-5} \cmidrule(lr){6-9}
         & SI-SNR & STOI  & WB-PESQ & NB-PESQ & SI-SNR & STOI  & WB-PESQ & NB-PESQ \\
    \midrule
    Noisy & 9.030  & 0.866  & 1.820  & 2.753  & 9.070  & 0.915  & 1.582  & 2.454 \\
    DCCRN-E\cite{hu20g_interspeech} & -     & -     & -     & 3.077  & -     & -     & -     & 3.266  \\
    PoCoNet\cite{isik20_interspeech} & -     & -     & 2.830  & -     & -     & -     & 2.750  & - \\
    DCCRN+\cite{lv21_interspeech} & -     & -     & -     & 3.300  & -     & -     & -     & 3.330  \\
    TRU-Net\cite{choi2021real} & 14.870  & 0.912  & 2.740  & 3.350  & 17.550  & 0.963  & 2.860  & 3.360  \\
    CTS-Net\cite{li2021two} & 15.580  & 0.930  & 3.020  & 3.470  & 17.990  & \textbf{0.967}  & \textbf{2.940}  & 3.420  \\
    FullSubNet\cite{hao2021fullsubnet} & 16.090  & 0.930  & 3.060  & \textbf{3.581}  & 17.440  & 0.961  & 2.810  & 3.403  \\
    \midrule
    CARNNHS-L(320, 160) & 13.808  & 0.908  & 2.316  & 2.701  & 15.364  & 0.957  & 2.034  & 2.616  \\
    CARNNHS-IL(320, 160) & 16.166  & 0.923  & 2.891  & 3.441  & 18.571  & 0.965  & 2.877  & 3.399  \\
    CARNNHS-L(512, 256) & 16.047  & 0.926  & 2.898  & 3.467  & 18.429  & 0.965  & 2.757  & 3.382  \\
    CARNNHS-IL(512, 256) & \textbf{16.704}  & \textbf{0.932}  & \textbf{3.063} & 3.519  & \textbf{18.803}  & \textbf{0.967}  & 2.892 & \textbf{3.431}  \\
    \bottomrule
    \end{tabular}%
  \label{tab:addlabel}%
\end{table*}%

The second part, i.e., the “RI+Mag” loss criterion is adopted to recover the complex spectrum as follows:
\begin{equation}
	\mathcal{L}_{\mathrm{mag}}=\frac{1}{T} \sum_t^T \sum_f^F \|S(t, f)|^p-|\hat{S}(t, f)|^p|^2 
\end{equation}
\begin{equation}
	\mathcal{L}_{\mathrm{RI}}=\frac{1}{T} \sum_t^T \sum_f^F \| S(t, f)|^p e^{j \theta_{S(t, f)}}-|\hat{S}(t, f)|^p e^{j \theta_{\hat{S}(t, f)}}|^2
\end{equation}
\begin{equation}
	\mathcal{L}=\mathcal{L}_{\mathrm{RI}}+\mathcal{L}_{\text {mag }}+\mathcal{L}_{\text {si-snr}},
\end{equation}
where $\mathcal{L}$ is the suggested method's loss function, and $p$ is a spectrum compression factor (set to 0.5). The operator $\theta$ determines the phase of a complex number.
\section{Experiments}
\label{sec:Experiments}

\subsection{Dataset}
\label{ssec:ataset}

A subset of the Interspeech 2021 DNS Challenge dataset was used to train and assess proposed method. The clean speech set consists of 562.72 hours of audio from 2150 speakers. The noise dataset consists of 181 hours of 60000 clips across 150 classes. During model training, we simulate speech-noise mixture as noisy speech via dynamic mixing \cite{hao2021fullsubnet}. Before the beginning of each training epoch, 75\% of the clean speechs are mixed with a randomly chosen room impulse response (RIR) from the openSLR26 and openSLR28 datasets \cite{ko2017study}. The speech-noise mixtures are then generated dynamically by combining clean speech and noise at a random SNR between -5 and 20 dB. The DNS Challenge presents a publicly accessible test dataset consisting of two classes of synthetic clips, namely those with and without reverberations. Each category contains 150 noise clips with an SNR ranging from 0 to 20 dB. This test set is used to evaluate the efficacy of the model.


\subsection{Experiments setup}
\label{ssec:Experiments setup}

For STFT, the window size is 32 ms, the shift is 16 ms, and the analysis is Hanning window.
We use 512-point DFT to extract 257-dimensional complex spectrum for 16 kHz sampling rate.
The model is optimized by Adam.
The initial learning rate is 0.001 and halved when the validation loss of two consecutive epochs no longer decreased. 
The batch size is 96.

The channel number of the convolutional layers in the encoder are \{16,32,43,86,172,172\}. The kernel size and the stride are respectively set to (3,3) and (1,2) in time and frequency dimensions. We apply zero-padding to the time direction but not to the frequency direction for all convolutions and deconvolutions. The hidden size of ARNBlock is set to 516.

\section{EXPERIMENT RESULTS AND ANALYSIS}
\label{sec:EXPERIMENT-RESULTS-AND-ANALYSIS}

The performance of each model on the DNS challenge test dataset is displayed in Table \ref{tab:addlabel}. "With Reverb" and "Without Reverb" refer to test sets with and without reverberation. The proposed intelligent neural homomorphic synthesis approach, called CARNNHS, employs two CARNs to estimate the excitation and the vocal tract. CARNNHS-L refers to the approach of segmenting excitation and the vocal tract via classic liftering. The excitation and vocal tract are segmented in CARNNHS-IL using the proposed intelligence liftering.

In Table \ref{tab:addlabel}, we contrast CARNNHS-IL with some recent SOTA approaches \cite{hu20g_interspeech,isik20_interspeech,lv21_interspeech,choi2021real,li2021two,hao2021fullsubnet}. Compared to the latest method, our proposed CARNNHS-IL shows better performance in noise reduction tests without reverberation and retains its advantages in the presence of reverberation. The proposed method significantly improves the performance of the SI-SNR measure. This demonstrates that the intelligent neural homomorphic synthesis speech enhancement method has a more extraordinary ability to recover time-domain waveforms.

To investigate the effectiveness of intelligent liftering, we present comparative experiment results in the last four rows of Table \ref{tab:addlabel}. Intelligent liftering is more robust than traditional liftering at two Fourier transform sizes. As proposed in \cite{jiang2022speech}, in order to alleviate cepstrum aliasing, the FFT size needs to be increased. This will undoubtedly increase the computational complexity. Cepstral aliasing will cause the traditional liftering unable to separate the excitation and the vocal tract. Nevertheless, our proposed intelligent liftering can alleviate the aliasing problem by adopting ratio masking. The experimental results in Table \ref{tab:addlabel} demonstrate that CARNNHS-IL can maintain performance even when the Fourier transform size is reduced. In contrast, the performance of CARNNHS-I drops significantly.
\begin{figure}[ht]
\centering
\subfigure[Traditional liftering for excitation]{ 
\begin{minipage}{3.5cm}
\centering 
\includegraphics[width=3.5cm]{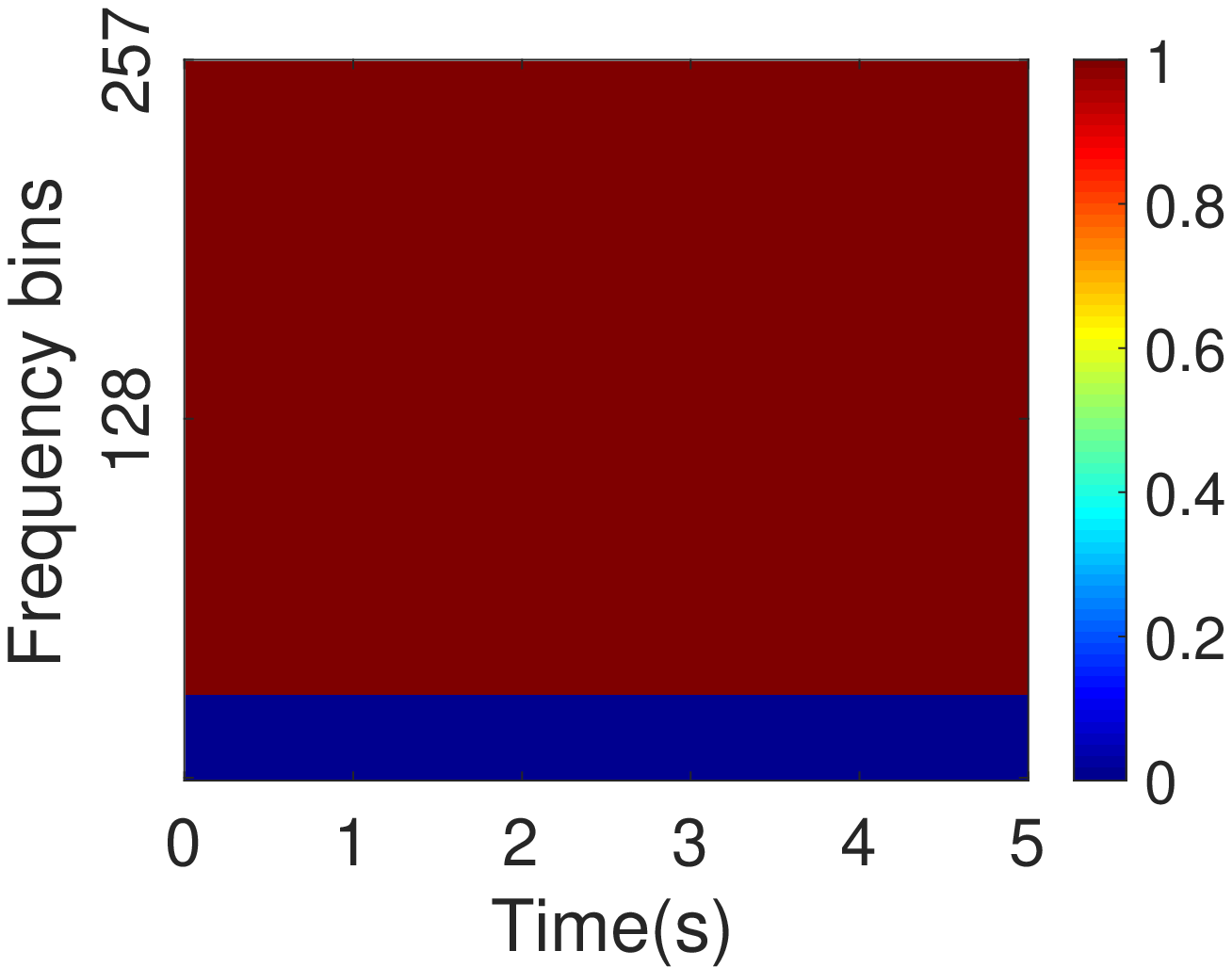}
\end{minipage}
}
\subfigure[Intelligent liftering for excitation]{ 
\begin{minipage}{3.5cm}
\centering 
\includegraphics[width=3.5cm]{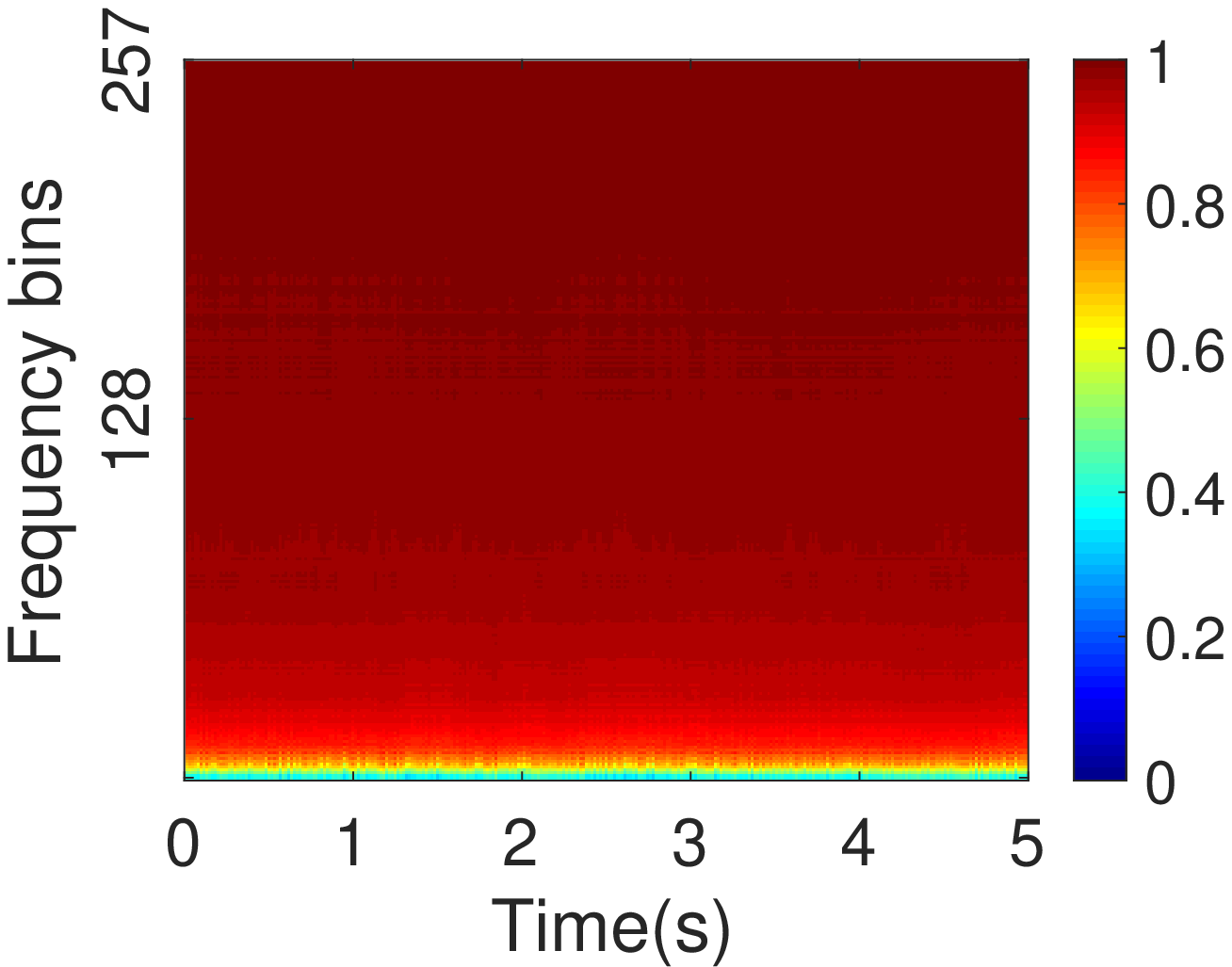}
\end{minipage}
}

\caption{A visualization of liftering and intelligent liftering. (a) A visualization of traditional liftering for excitation. (b) A visualization of intelligent liftering for excitation.}
\label{fig:Result}
\end{figure}

Fig. \ref{fig:Result} illustrates the visualization of traditional liftering and intelligent liftering for excitation, respectively. As mentioned in 2.3, although intelligent liftering does not have a clear target, since the dynamic range of excitation and vocal tract in the cepstral domain is significantly different, the cepstral domain ratio mask predicted by ARN can naturally segment the two.

\section{CONCLUSIONS}
\label{sec:CONCLUSIONS}
This paper proposes some practical improvements to the speech enhancement method of neural homomorphic synthesis. We propose an intelligent liftering for more explicit segmentation of excitations and vocal tracts in neural homomorphic signal processing. Compared with the traditional neural homomorphic signal processing process, our proposed method has stronger performance, overcomes the cepstral aliasing problem, and is more suitable for the case of a small Fourier transform size. In the next step, we will look at how different structures handle excitation and vocal tract separately.

\textbf{Acknowledgments}: This research was partly supported by the China National Nature Science Foundation (No. 61876214).

\newpage
\bibliographystyle{IEEEbib}
\bibliography{strings,refs}

\begin{thebibliography}{10}

\bibitem{boll1979suppression}
Steven Boll,
\newblock ``Suppression of acoustic noise in speech using spectral
  subtraction,''
\newblock {\em IEEE Transactions on acoustics, speech, and signal processing},
  vol. 27, no. 2, pp. 113--120, 1979.

\bibitem{lim1979enhancement}
Jae~Soo Lim and Alan~V Oppenheim,
\newblock ``Enhancement and bandwidth compression of noisy speech,''
\newblock {\em Proceedings of the IEEE}, vol. 67, no. 12, pp. 1586--1604, 1979.

\bibitem{ephraim1984speech}
Yariv Ephraim and David Malah,
\newblock ``Speech enhancement using a minimum-mean square error short-time
  spectral amplitude estimator,''
\newblock {\em IEEE Transactions on acoustics, speech, and signal processing},
  vol. 32, no. 6, pp. 1109--1121, 1984.

\bibitem{wang2013towards}
Yuxuan Wang and DeLiang Wang,
\newblock ``Towards scaling up classification-based speech separation,''
\newblock {\em IEEE Transactions on Audio, Speech, and Language Processing},
  vol. 21, no. 7, pp. 1381--1390, 2013.

\bibitem{wang2022attention}
Heming Wang, Xueliang Zhang, and DeLiang Wang,
\newblock ``Attention-based fusion for bone-conducted and air-conducted speech
  enhancement in the complex domain,''
\newblock in {\em ICASSP 2022-2022 IEEE International Conference on Acoustics,
  Speech and Signal Processing (ICASSP)}. IEEE, 2022, pp. 7757--7761.

\bibitem{liu21f_interspeech}
Jinjiang Liu and Xueliang Zhang,
\newblock ``{Inplace Gated Convolutional Recurrent Neural Network for
  Dual-Channel Speech Enhancement},''
\newblock in {\em Proc. Interspeech 2021}, 2021, pp. 1852--1856.

\bibitem{tan2018convolutional}
Ke~Tan and DeLiang Wang,
\newblock ``A convolutional recurrent neural network for real-time speech
  enhancement.,''
\newblock in {\em Interspeech}, 2018, pp. 3229--3233.

\bibitem{li2022filtering}
Andong Li, Chengshi Zheng, Guochen Yu, Juanjuan Cai, and Xiaodong Li,
\newblock ``Filtering and refining: A collaborative-style framework for
  single-channel speech enhancement,''
\newblock {\em IEEE/ACM Transactions on Audio, Speech, and Language
  Processing}, vol. 30, pp. 2156--2172, 2022.

\bibitem{wang2018supervised}
DeLiang Wang and Jitong Chen,
\newblock ``Supervised speech separation based on deep learning: An overview,''
\newblock {\em IEEE/ACM Transactions on Audio, Speech, and Language
  Processing}, vol. 26, no. 10, pp. 1702--1726, 2018.

\bibitem{mowlaee12_interspeech}
Pejman Mowlaee, Rahim Saeidi, and Rainer Martin,
\newblock ``{Phase estimation for signal reconstruction in single-channel
  source separation},''
\newblock in {\em Proc. Interspeech 2012}, 2012, pp. 1548--1551.

\bibitem{pandey2022self}
Ashutosh Pandey and DeLiang Wang,
\newblock ``Self-attending rnn for speech enhancement to improve cross-corpus
  generalization,''
\newblock {\em IEEE/ACM Transactions on Audio, Speech, and Language
  Processing}, vol. 30, pp. 1374--1385, 2022.

\bibitem{pandey2019tcnn}
Ashutosh Pandey and DeLiang Wang,
\newblock ``Tcnn: Temporal convolutional neural network for real-time speech
  enhancement in the time domain,''
\newblock in {\em ICASSP 2019-2019 IEEE International Conference on Acoustics,
  Speech and Signal Processing (ICASSP)}. IEEE, 2019, pp. 6875--6879.

\bibitem{pandey2022tparn}
Ashutosh Pandey, Buye Xu, Anurag Kumar, Jacob Donley, Paul Calamia, and DeLiang
  Wang,
\newblock ``Tparn: Triple-path attentive recurrent network for time-domain
  multichannel speech enhancement,''
\newblock in {\em ICASSP 2022-2022 IEEE International Conference on Acoustics,
  Speech and Signal Processing (ICASSP)}. IEEE, 2022, pp. 6497--6501.

\bibitem{li2022use}
Kai Li, Xiaolin Hu, and Yi~Luo,
\newblock ``On the use of deep mask estimation module for neural source
  separation systems,''
\newblock {\em arXiv preprint arXiv:2206.07347}, 2022.

\bibitem{luo2020dual}
Yi~Luo, Zhuo Chen, and Takuya Yoshioka,
\newblock ``Dual-path rnn: efficient long sequence modeling for time-domain
  single-channel speech separation,''
\newblock in {\em ICASSP 2020-2020 IEEE International Conference on Acoustics,
  Speech and Signal Processing (ICASSP)}. IEEE, 2020, pp. 46--50.

\bibitem{luo2019conv}
Yi~Luo and Nima Mesgarani,
\newblock ``Conv-tasnet: Surpassing ideal time--frequency magnitude masking for
  speech separation,''
\newblock {\em IEEE/ACM transactions on audio, speech, and language
  processing}, vol. 27, no. 8, pp. 1256--1266, 2019.

\bibitem{ge2021multi}
Meng Ge, Chenglin Xu, Longbiao Wang, Eng~Siong Chng, Jianwu Dang, and Haizhou
  Li,
\newblock ``Multi-stage speaker extraction with utterance and frame-level
  reference signals,''
\newblock in {\em ICASSP 2021-2021 IEEE International Conference on Acoustics,
  Speech and Signal Processing (ICASSP)}. IEEE, 2021, pp. 6109--6113.

\bibitem{du2020self}
Zhihao Du, Ming Lei, Jiqing Han, and Shiliang Zhang,
\newblock ``Self-supervised adversarial multi-task learning for vocoder-based
  monaural speech enhancement.,''
\newblock in {\em INTERSPEECH}, 2020, pp. 3271--3275.

\bibitem{du2020joint}
Zhihao Du, Xueliang Zhang, and Jiqing Han,
\newblock ``A joint framework of denoising autoencoder and generative vocoder
  for monaural speech enhancement,''
\newblock {\em IEEE/ACM Transactions on Audio, Speech, and Language
  Processing}, vol. 28, pp. 1493--1505, 2020.

\bibitem{liu2020neural}
Zhijun Liu, Kuan Chen, and Kai Yu,
\newblock ``Neural homomorphic vocoder.,''
\newblock in {\em INTERSPEECH}, 2020, pp. 240--244.

\bibitem{jiang2022speech}
Wenbin Jiang, Zhijun Liu, Kai Yu, and Fei Wen,
\newblock ``Speech enhancement with neural homomorphic synthesis,''
\newblock in {\em ICASSP 2022-2022 IEEE International Conference on Acoustics,
  Speech and Signal Processing (ICASSP)}. IEEE, 2022, pp. 376--380.

\bibitem{rabiner2010theory}
Lawrence Rabiner and Ronald Schafer,
\newblock {\em Theory and applications of digital speech processing},
\newblock Prentice Hall Press, 2010.

\bibitem{hu20g_interspeech}
Yanxin Hu, Yun Liu, Shubo Lv, Mengtao Xing, Shimin Zhang, Yihui Fu, Jian Wu,
  Bihong Zhang, and Lei Xie,
\newblock ``{DCCRN: Deep Complex Convolution Recurrent Network for Phase-Aware
  Speech Enhancement},''
\newblock in {\em Proc. Interspeech 2020}, 2020, pp. 2472--2476.

\bibitem{isik20_interspeech}
Umut Isik, Ritwik Giri, Neerad Phansalkar, Jean-Marc Valin, Karim Helwani, and
  Arvindh Krishnaswamy,
\newblock ``{PoCoNet: Better Speech Enhancement with Frequency-Positional
  Embeddings, Semi-Supervised Conversational Data, and Biased Loss},''
\newblock in {\em Proc. Interspeech 2020}, 2020, pp. 2487--2491.

\bibitem{lv21_interspeech}
Shubo Lv, Yanxin Hu, Shimin Zhang, and Lei Xie,
\newblock ``{DCCRN+: Channel-Wise Subband DCCRN with SNR Estimation for Speech
  Enhancement},''
\newblock in {\em Proc. Interspeech 2021}, 2021, pp. 2816--2820.

\bibitem{choi2021real}
Hyeong-Seok Choi, Sungjin Park, Jie~Hwan Lee, Hoon Heo, Dongsuk Jeon, and Kyogu
  Lee,
\newblock ``Real-time denoising and dereverberation wtih tiny recurrent
  u-net,''
\newblock in {\em ICASSP 2021-2021 IEEE International Conference on Acoustics,
  Speech and Signal Processing (ICASSP)}. IEEE, 2021, pp. 5789--5793.

\bibitem{li2021two}
Andong Li, Wenzhe Liu, Chengshi Zheng, Cunhang Fan, and Xiaodong Li,
\newblock ``Two heads are better than one: A two-stage complex spectral mapping
  approach for monaural speech enhancement,''
\newblock {\em IEEE/ACM Transactions on Audio, Speech, and Language
  Processing}, vol. 29, pp. 1829--1843, 2021.

\bibitem{hao2021fullsubnet}
Xiang Hao, Xiangdong Su, Radu Horaud, and Xiaofei Li,
\newblock ``Fullsubnet: A full-band and sub-band fusion model for real-time
  single-channel speech enhancement,''
\newblock in {\em ICASSP 2021-2021 IEEE International Conference on Acoustics,
  Speech and Signal Processing (ICASSP)}. IEEE, 2021, pp. 6633--6637.

\bibitem{ko2017study}
Tom Ko, Vijayaditya Peddinti, Daniel Povey, Michael~L Seltzer, and Sanjeev
  Khudanpur,
\newblock ``A study on data augmentation of reverberant speech for robust
  speech recognition,''
\newblock in {\em 2017 IEEE International Conference on Acoustics, Speech and
  Signal Processing (ICASSP)}. IEEE, 2017, pp. 5220--5224.

\end{thebibliography}

\end{document}